\documentclass[10pt,journal,cspaper,compsoc]{IEEEtran}
\usepackage{eurosym}
\usepackage{color}
\usepackage{refcount}

\usepackage{varioref}
\usepackage{multirow}
\usepackage[normalem]{ulem}
\usepackage{cite}
\ifCLASSOPTIONcompsoc
\else

\fi
\usepackage[pdftex]{graphicx}
\DeclareGraphicsExtensions{.pdf, .eps}
\ifCLASSINFOpdf
\else
\fi
\usepackage[cmex10]{amsmath}
\usepackage{array}

\usepackage{amssymb}
\begin{document}

\title{Critical Study of Markovian Approaches for Batch Arrival Modeling in IEEE~802.15.4-based Networks}

\author{Mohammad Sayad Haghighi
\thanks{M. Sayad Haghighi is now with the School of Electrical and Computer Engineering, University of Tehran, Iran. Email: sayad [at] ieee.org } \thanks{Technical Report No. 1204-495-2019}\thanks{The contents are mainly from our technical report released in 2013.}
}

\IEEEcompsoctitleabstractindextext{%
\begin{abstract}
    The transient non-stationary nature of network reaction to batch arrivals has led to complex models in order to estimate the network quantities of interest. In this report we focus on the IEEE 802.15.4 standard MAC layer and try to deeply study the current approaches in modeling the network reaction to one-shot data arrivals. In addition to the general description of models, we mention the positive and negative points in each case proposing possible alternations which could improve the accuracy of the models. 
\end{abstract}
}
\maketitle

\section{Introduction}
\begin{figure}[t]
\begin{center}
\includegraphics[width=150pt]{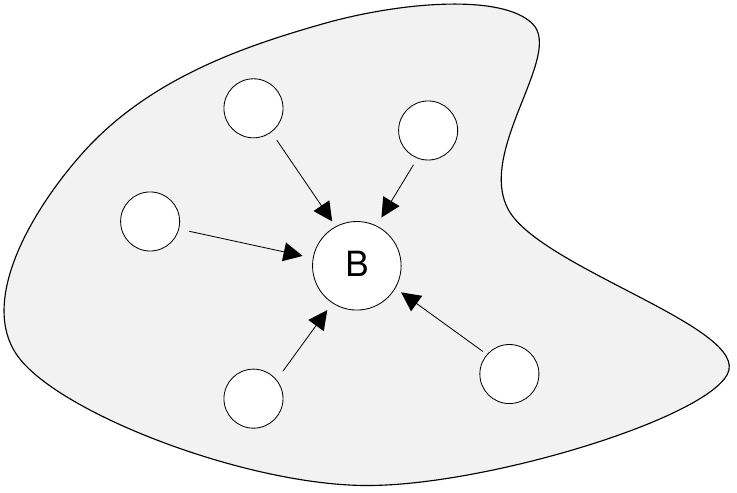}   
\caption{Batch arrival demonstration: a set of stations receive packets at their MAC layer in order to be dispatched to a single destination.}
\label{fig:batcharrival}
\end{center}
\end{figure}

Instances of batch data arrival can be seen in many applications. For example in sensor networks an event occurrence might trigger many sensors to send the sensed values to the cluster head (base station) located in their vicinity.  Alternatively, a station may issue a request through broadcasting to collect some local information \cite{haghighi2008securing}. For example, in routing protocols, building the neighbors list table is vital for every station participating in the route discovery. Therefore, it is necessary to build/update the table every so often. This is usually accomplished through broadcasting a request called Hello. Every station receiving this request sends its ID back to the sender. In general, multiple stations receive the request simultaneously due to the nature of request broadcasting. This leads to the batch arrival of data to the MAC layers of responding stations. A demonstration of batch arrival is depicted in Fig.~\ref{fig:batcharrival}.
The major quantities of interest in the analysis of such scenarios are usually the probability of success in data delivery
for the responding stations, the collision rate and the mean and sometimes the distribution of delay for the collection of responses.

In this report we review the current models for the analysis of batch arrivals with IEEE 802.15.4 as the MAC layer. In each case we do a critical revision and propose possible alternations and manipulations to the model which could increase the accuracy of predictions.
\section{IEEE 802.15.4 MAC protocol specifications}

The IEEE 802.15.4 standard \cite{IEEE802154standard} defines two operational mode for the MAC layer; Beacon-enabled and Non-beacon-enabled. In the former one, the cluster head (Coordinator) periodically advertises a beacon with which the super-frame starts. In beaconless mode, the cluster head does not broadcast beacon messages and the network operates asynchronously. All of the current analyses have focused on the beacon-enabled mode and its slotted CSMA/CA contention mechanism.
This is due to the nature of batch arrival problem in which there is an inherent synchronization point; either a broadcasted request or a general event. Markov models have proven to be useful tools in the analysis of network performance \cite{toorchi2013markov}. 
The Markov model of the slotted CSMA/CA protocol used in the IEEE 802.15.4 standard is depicted in Fig.~2.

Each station which has a packet to sent starts waiting for a random number time-slots chosen from the interval $[0,2^{BE_{min}}]$. Then it enters the Clear Channel Assessment (CCA) state and senses the channel. If the channel is busy it defers its transmission for a random number of time-slots selected from the interval $[0,2^{min(BE_{min}+k,BE_{max})}]$, where $k$ is the number of times the station has already postponed its transmission. If the channel is continuously sensed free for $CW$   time-slots, the station starts transmitting over the channel.
At the other side, if the MAC layer cannot capture the channel after $NB{max}$ tries, it gives up trying and reports failure to the upper layer. The standard has defined $CW=2$, $BE_{min}$=3, $BE_{max}=5$ and $NB_{max}=4 $ as the default values. 

The transmission of acknowledgements in response to packet reception is optional in the standard. As we will see, many of the current models assume working in the NACK
mode. Aside from the analysis simplification reasons, this is usually do to the assumption of the existence of redundancy in the responses. While this assumption holds in applications like data collection in sensor networks, it is not always true. According to \cite{Ramachandran2007}, setting CW to 2 in the NACK mode is not necessary. Thus for energy conservation purposes and to increase in transmission efficiency, it is usually assumed that CW=1 in the NACK mode. 
\section{Classification of the Models}
\subsection{Markov-Model-based Approaches}

\begin{figure}[b]
\begin{center}
  \includegraphics[width=\columnwidth]{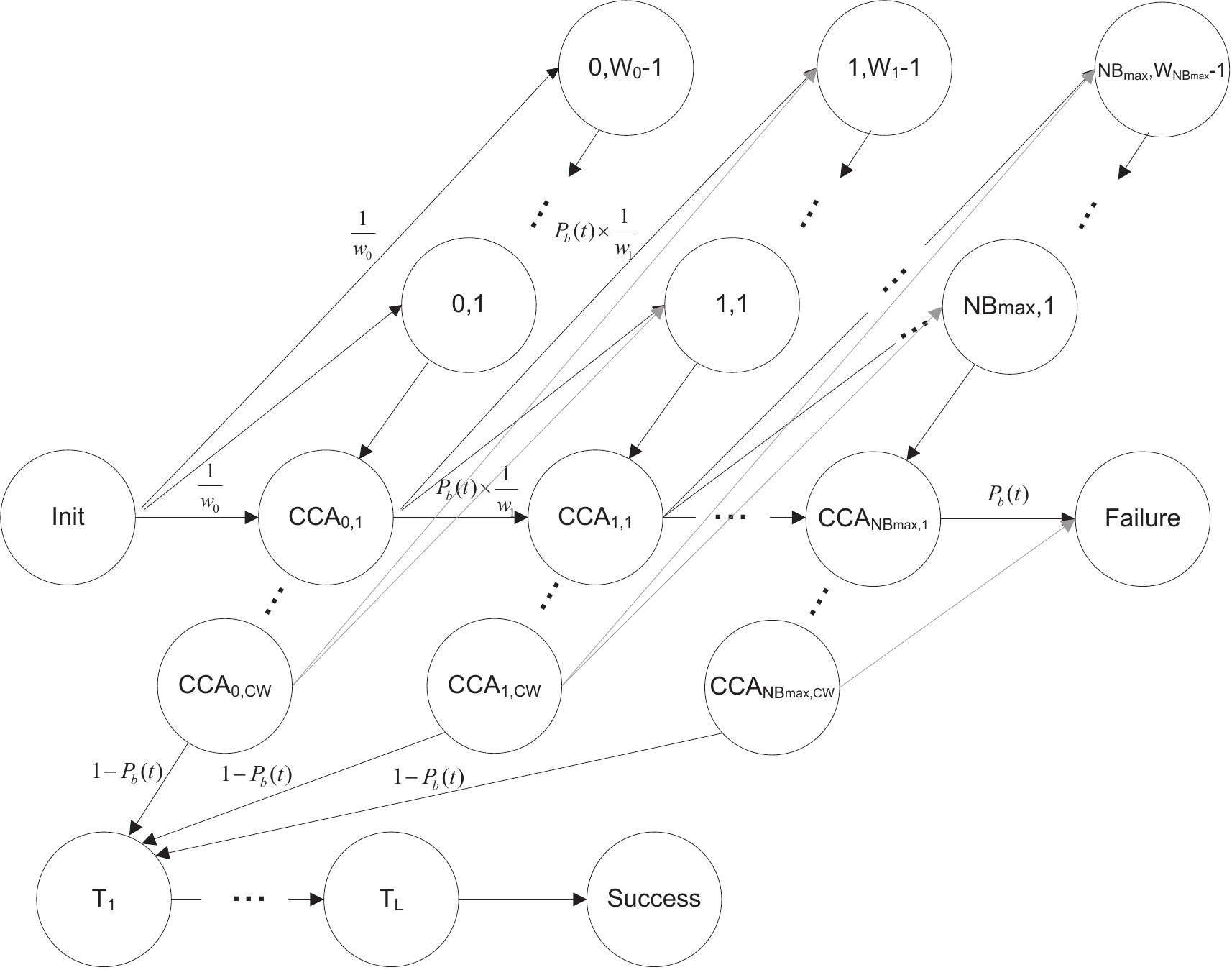}
    \caption{Markov model of the (slotted) MAC layer of the IEEE 802.15.4 standard. }
\end{center}
  \label{IEEE802154figure}
\end{figure}

 Almost all of the Markov-model-based approaches rely on the notion of attempt probability which was initially used in the analysis of truncated binary exponential back-off algorithm for Ethernet
\cite{molle1994new}.
Having the binary exponential mechanism of Fig. \ref{IEEE802154figure} in mind, we may define the attempt probability as follows:
\begin{equation}
a(t)=\sum^{NB_{max}}_{k=0}d_k(t)
\end{equation}
where:
\begin{equation}
d_k(t)=b_0(t)\circledast b_1(t)~\ldots\circledast b_k(t)
\end{equation}   
\begin{equation}
b_i(t)=\begin{cases}\frac{1}{W_i} & 0\leqslant t<W_i \\
0 & oth. \\
\end{cases}~;~W_i=2^{min(BE_{min}+i,BE_{max})} 
\end{equation} 

Notice that there is no concept of the channel in the attempt probability definition. It solely represents the probability of reaching time-slot $t$ with any combination of allowable jumps (back-offs).
This independency of the attempt probability from the channel status is valuable since it can mitigate the complication in the analysis of stations dependent transmission behavior. There is a single channel   and the stations are contending with each other to take over it and this makes the transmissions dependent. 

\subsubsection{Leibnitz et al.'s Model}
\begin{figure*}[!htb]
\begin{center}
 \includegraphics[width=\textwidth]{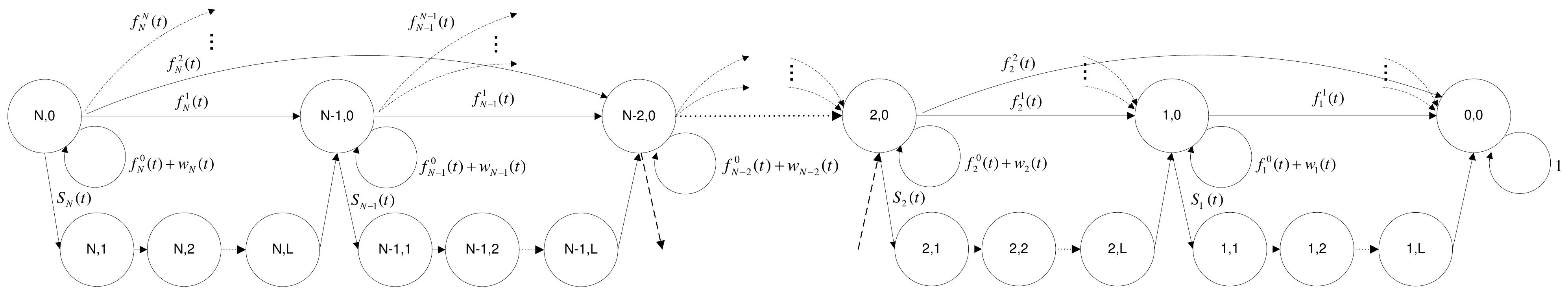}
    \caption{The state space of Leibnitz et al.'s model with time-dependent transition probabilities.  } 
\label{leibnitz1}
\end{center}
\end{figure*}

\paragraph{Model Description}
  Leibnitz et al. \cite{leibnitz2005} were the first who adopted the idea of using attempt probability for modeling the IEEE 802.15.4 MAC protocol under batch arrival traffic assumption. They used the Markov model depicted in Fig.~\ref{leibnitz1} to follow the global state of the system as the time goes by. 

In this figure, the number of stations is represented by $N$ and $L$ is the packet length in slots. Assuming to be at time-slot $t$, 
$s_i(t)$ is the probability that only one of the $i$ remaining stations attempts to access the channel at time-slot $t$. We will refer to the term "access" later in the discussion part. $w_i(t)$ is the probability that none of the $i$ remaining stations attempts to access the channel,  and, $f_i^j(t)$ is the probability that $j$ of the $i$ remaining stations do their $NB_{max}$ attempt at time-slot $t$ and fail to find the channel free.
The following equations summarizes the Leibnitz et al.'s model:

\begin{equation}
s_i(t)=\begin{pmatrix}i \\
1 \\
\end{pmatrix}a(t)(1-a(t))^{i-1}
\label{equation:sit}
\end{equation}
\begin{equation}
w_i(t)=\begin{pmatrix}i \\
0 \\
\end{pmatrix}(1-a(t))^{i}
\end{equation}    
\label{equation:wit}
\begin{align}
&f_i^j(t)=c_i(t)\begin{pmatrix}i \\
j\end{pmatrix}\eta(t)^j(1-\eta(t))^{i-j}~~~\\&~~~c_i(t)=1-s_i(t)-w_i(t) ~~;~~ \eta(t)=d_{NB_{max}}(t).\xi
\label{equation:six}
\end{align}
Notice that in the above equations, $c_i(t)$ is actually the probability of collision. The channel status is reflected in the channel busy probability which is assumed to be a constant known as $\xi$ in the Leibnitz et al.'s model and is given by Eq. (\ref{equation:Leibnitz_zetha}). Notice that $E[W]$ is the mean back-off window size.
\begin{equation}
\xi=min\{ 1,\frac{L(N-1)}{E\{W\}}\}
\label{equation:Leibnitz_zetha}
\end{equation}
If $x_{i,j}(t_{max})$ refers to the probability of being in state $(i,j)$ at the maximum possible reachable time-slot by the binary exponential back-offs with a given set of timing parameters $\{BE_{min},BE_{max}, NB_{max}\}$, Leibnitz et al. claim one can find the number of successful transmissions using the following equation based on a modified states space of the model shown in Fig. \ref{leibnitz1} where we only distinguish between the transitions for successful and unsuccessful attempts. However it is unclear that how this modified state space is built up.
\begin{equation}
S_N=\sum_{i=0}^N[(N-i)\sum_{j=0}^Lx_{i,j}(t_{max})]
\label{equation:Leibnitz_SN}
\end{equation}

\paragraph{Criticizing the Model}
Although  Leibnitz et al.'s model is very inspiring, it suffers from severe flaws. The fixed channel utilization assumption is quite unrealistic since the number of nodes contending over the channel is expected to decreases over time.
Moreover, according to \cite{molle1994new}, $a(t)$ is probability that a given station attempts a transmission at the $t^{th}$ time-slot, \textit{assuming none of its earlier attempts were successful}. In the IEEE 802.15.4 protocol, due to the existence of channel assessments, a node senses the channel in the CCA state(s) before starting the transmission  (See Fig. \ref{IEEE802154figure}). Therefore, here the term "unsuccessful" shall refer to the event that the channel has been busy in the previous attempts. However, the existence of the term $f_i^0(t)$ in the model indicates that the authors have assumed some kind of retransmission policy when collisions occur. The IEEE 802.15.4 standard \cite{IEEE802154standard} does not explicitly define any retransmission policy upon detection of collision. Thus, this must be a vendor-dependent implementation factor. The detection of collisions in  Collision-Avoiding (CA) wireless medium access control protocols is usually done using the time-outs for the reception of acknowledgement packets.
Fig. \ref{leibnitz1} indicates that if the system is in state $(i,0)$ and a station faces collision at time-slot $t$, it will remain in the same state at $t+1$, while in a real scenario the detection of collision takes at least $L$ time-slots to which we have to add the ACK packet delivery time. Aside from this, the ACK is sent over the same common channel in the IEEE 802.15.4 standard and usually lasts one time-slot. The standard has set a contention window size of two to avoid collisions on acknowledgement packets. 
The model also implies a specific retransmission policy in which every colliding station continues doing the binary exponential back-offs without resetting any of the MAC parameters. So according to the Markov model Leibnitz et al. presented, collisions are treated just like the busy channel assessments. 

\begin{figure*}[!ht]
\includegraphics[width=\textwidth]{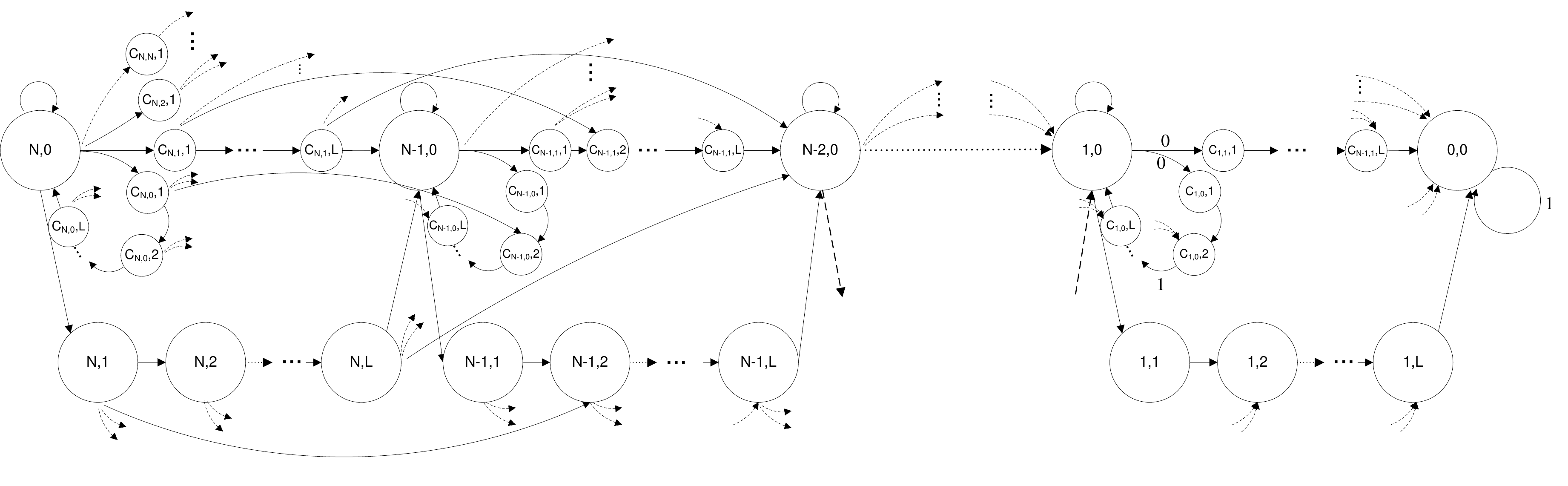}
\caption{The improved version of Leibnitz et al.'s Markov model}
\label{figure:leibnitz_improved}
\end{figure*}

Another problematic aspect of the model is the inability of counting the failures in the transition states which account for the successful transmissions. The model only allows failures to occur while the system is in state $(i,0)$ which implies there exists no ongoing transmission. When there exists no active transmission over the channel, it must be logically free. Therefore, there can exist only two possibilities: collision, or, a successful transmission. Calculation of the failure probability using Eq. (\ref{equation:six})
while the channel is free is wrong. The existence of the term $\xi$ implies that the authors have assumed the channel can be busy at $t$. So a better Markov model would be like the one shown in Fig. \ref{figure:leibnitz_improved}. In this model, the failures can occur only in the transition states i.e.
when the channel is busy. Moreover, 
if a collision happens, the channel remains busy for the entire packet length.
So, unlike the original model, when a collision occurs in state $(i,0)$, it does not simply move on to the $(j,0)~;~j\leqslant{i}$. Notice that we have not changed the retransmission policy in the modified model.

Although Leibnitz et al. claim that the number of successful transmissions can be found using Eq.~(\ref{equation:Leibnitz_SN}) from a modified state space, they have not clearly defined what exactly the modified state space is. Clearly applying Eq.~(\ref{equation:Leibnitz_SN}) to the original model of Fig.~\ref{leibnitz1} is wrong as being in state ${i,j}$ at $t_{max}$ does not imply that we have had $N-i$ successful transmissions. A better approach which is also applicable to the improved model shown in Fig.~\ref{figure:leibnitz_improved}, is to take advantage of the renewal theory which leads to the following formula for calculation of the number of successful transmissions. 

\begin{equation}
S_N=\sum_{t=1}^{t_{max}}\sum_{i=1}^N x_{i,1}(t)
\end{equation}
where unlike Eq.~(\ref{equation:six}), $x_{i,1}(t)$ is obtained from the original state space.

Another issue with Leibnitz et al.'s model is the misunderstanding of CCA mechanism. This problem can also be found in the other Markov-model-based approaches came after \cite{shu2007packetloss,li2009analytical}. According to the IEEE 802.15.4 standard,  if a station senses the channel busy in CCA state, it will choose a random number in the interval $[0,2^{min(BE_{min}+k,BE_{max})}]$ at the $k^{th}$ back-off ($k\leqslant{NB_{max}})$ and waits for that number of time-slots before sensing the channel again. The choice of $0$ in the above interval means that the station does not wait at all and sense the channel immediately in the \textit{next} time-slot (see Fig.~\ref{IEEE802154figure}). By the definition of attempt probability, the authors have assumed that the choice of $0$ means sensing the same time-slot which was already sensed once which is meaningless. Therefore, both $t_{max}$ value and the attempt probability equation  have to be corrected.  The correct value of $t_{max}$ is given by Eq.~\ref{equation:t_max}.
\begin{align}
t_{max}&=\sum_{j=BE_{min}}^{BE_{max}}2^j+(NB_{max}-(BE_{max}-BE_{min}))\notag\\&\times2^{BE_{max}}-1\\
&=(2+NB_{max}-(BE_{max}-BE_{min}) )\notag\\&\times2^{BE_{max}}-2^{BE_{min}}-1
\label{equation:t_max}
\end{align}

We also have to 
remark on Fig. 5a of Leibnitz et al.'s paper. In reality, since the authors treat the collisions just like busy channel assessments, a station
has to either transmit or abandon trying to transmit because of failure before
reaching $t_{max}$. However, since the model works based on the attempt probability notion which is independent of the channel and stations state, it will force the Markov model to enter a dead-lock status as $t$ approaches $t_{max}$ which does not necessarily guarantee that every node reaches state $(0,0)$. For example, assume that the model is in state (2,0) at $t_{max}-1$. We expect that in the next time-slot, the two remaining stations commit a simultaneous transmission as the binary exponential back-off mechanism enforces this. But, taking a closer look at equations (\ref{equation:sit}) through (\ref{equation:six}), reveals a controversial  fact. Since $a(t)$ is very small at $t_{max}-1$, $w_i(t)$ gets close to one and $s_i(t)$ becomes almost zero. Thus, $c_i(t)$ also approaches zero and the Model tends to stay at the same state instead of moving to $(0,0)$. In fact, this problem is more general. At the ending time-slots, almost all of $(i,0)$ states ($i\neq0$) turn absorbent.   

From the above discussion it is clear that the model does not necessarily end up reaching state $(0,0)$ at $t_{max}$ specially when $N$ is large. So, $x_{0,0}(t)$ does not represent the cumulative probability of $\tau_{head}(t)$ which is the probability mass function of the total transmission delay, i.e. the time required for all the cluster members to either transmit their data to the cluster head or fail. 


\section{Final Remarks}
We have conducted a critical approach to analyse the weaknesses of Markov-based approaches in modeling the batch traffic  characteristics in IEEE\ 802.15.4 networks. More specifically, we focused on a Markov-based model and discussed the flaws in it giving some hints for mitigation of those flaws. As of today, the research in this area has advanced further and very precise Markovian models have been introduced. The readers are encouraged to read our works in \cite{haghighi2011stochastic} and~\cite{haghighi2014stochastic}. 
\bibliographystyle{IEEEtran}
\bibliography{references}

\end{document}